%% file: PerugiaProceeding.tex
\documentclass[12pt]{article}
\usepackage{graphicx}
\usepackage{epstopdf}
\usepackage{subfig}
\usepackage{caption}
\usepackage{mathrsfs}
\usepackage{amssymb}
\usepackage{amsmath}
\usepackage{verbatim}
\advance\textheight by \topskip
\begin{document}


\newcommand{\HPA}[1]{{\it Helv.\ Phys.\ Acta.\ }{\bf #1}}
\newcommand{\AP}[1]{{\it Ann.\ Phys.\ }{\bf #1}}
\newcommand{\be}{\begin{equation}}
\newcommand{\ee}{\end{equation}}
\newcommand{\br}{\begin{eqnarray}}
\newcommand{\er}{\end{eqnarray}}
\newcommand{\ba}{\begin{array}}
\newcommand{\ea}{\end{array}}
\newcommand{\bi}{\begin{itemize}}
\newcommand{\ei}{\end{itemize}}
\newcommand{\bn}{\begin{enumerate}}
\newcommand{\en}{\end{enumerate}}
\newcommand{\bc}{\begin{center}}
\newcommand{\ec}{\end{center}}
\newcommand{\ul}{\underline}
\newcommand{\ol}{\overline}
\def\l{\left\langle}
\def\r{\right\rangle}
\def\as{\alpha_{s}}
\def\ycut{y_{\mbox{\tiny cut}}}
\def\yij{y_{ij}}
\def\epem{\ifmmode{e^+ e^-} \else{$e^+ e^-$} \fi}
\newcommand{\eeww}{$e^+e^-\rightarrow W^+ W^-$}
\newcommand{\qqQQ}{$q_1\bar q_2 Q_3\bar Q_4$}
\newcommand{\eeqqQQ}{$e^+e^-\rightarrow q_1\bar q_2 Q_3\bar Q_4$}
\newcommand{\eewwqqqq}{$e^+e^-\rightarrow W^+ W^-\ar q\bar q Q\bar Q$}
\newcommand{\eeqqgg}{$e^+e^-\rightarrow q\bar q gg$}
\newcommand{\eeqloop}{$e^+e^-\rightarrow q\bar q gg$ via loop of quarks}
\newcommand{\eeqqqq}{$e^+e^-\rightarrow q\bar q Q\bar Q$}
\newcommand{\eewwjjjj}{$e^+e^-\rightarrow W^+ W^-\rightarrow 4~{\rm{jet}}$}
\newcommand{\eeqqggjjjj}{$e^+e^-\rightarrow q\bar 
q gg\rightarrow 4~{\rm{jet}}$}
\newcommand{\eeqloopjjjj}{$e^+e^-\rightarrow q\bar 
q gg\rightarrow 4~{\rm{jet}}$ via loop of quarks}
\newcommand{\eeqqqqjjjj}{$e^+e^-\rightarrow q\bar q Q\bar Q\rightarrow
4~{\rm{jet}}$}
\newcommand{\eejjjj}{$e^+e^-\rightarrow 4~{\rm{jet}}$}
\newcommand{\jjjj}{$4~{\rm{jet}}$}
\newcommand{\qqbar}{$q\bar q$}
\newcommand{\ww}{$W^+W^-$}
\newcommand{\ar}{\rightarrow}
\newcommand{\sm}{${\cal {SM}}$}
\newcommand{\Dir}{\kern -6.4pt\Big{/}}
\newcommand{\Dirin}{\kern -10.4pt\Big{/}\kern 4.4pt}
\newcommand{\DDir}{\kern -8.0pt\Big{/}}
\newcommand{\DGir}{\kern -6.0pt\Big{/}}
\newcommand{\wwqqqq}{$W^+ W^-\ar q\bar q Q\bar Q$}
\newcommand{\qqgg}{$q\bar q gg$}
\newcommand{\qloop}{$q\bar q gg$ via loop of quarks}
\newcommand{\qqqq}{$q\bar q Q\bar Q$}

\def\st{\sigma_{\mbox{\scriptsize t}}}
\def\Ord{\buildrel{\scriptscriptstyle <}\over{\scriptscriptstyle\sim}}
\def\OOrd{\buildrel{\scriptscriptstyle >}\over{\scriptscriptstyle\sim}}
\def\jhep #1 #2 #3 {{JHEP} {\bf#1} (#2) #3}
\def\plb #1 #2 #3 {{Phys.~Lett.} {\bf B#1} (#2) #3}
\def\npb #1 #2 #3 {{Nucl.~Phys.} {\bf B#1} (#2) #3}
\def\epjc #1 #2 #3 {{Eur.~Phys.~J.} {\bf C#1} (#2) #3}
\def\zpc #1 #2 #3 {{Z.~Phys.} {\bf C#1} (#2) #3}
\def\jpg #1 #2 #3 {{J.~Phys.} {\bf G#1} (#2) #3}
\def\prd #1 #2 #3 {{Phys.~Rev.} {\bf D#1} (#2) #3}
\def\prep #1 #2 #3 {{Phys.~Rep.} {\bf#1} (#2) #3}
\def\prl #1 #2 #3 {{Phys.~Rev.~Lett.} {\bf#1} (#2) #3}
\def\mpl #1 #2 #3 {{Mod.~Phys.~Lett.} {\bf#1} (#2) #3}
\def\rmp #1 #2 #3 {{Rev. Mod. Phys.} {\bf#1} (#2) #3}
\def\cpc #1 #2 #3 {{Comp. Phys. Commun.} {\bf#1} (#2) #3}
\def\sjnp #1 #2 #3 {{Sov. J. Nucl. Phys.} {\bf#1} (#2) #3}
\def\xx #1 #2 #3 {{\bf#1}, (#2) #3}
\def\hepph #1 {{\tt hep-ph/#1}}
\def\preprint{{preprint}}

\def\beq{\begin{equation}}
\def\beeq{\begin{eqnarray}}
\def\eeq{\end{equation}}
\def\eeeq{\end{eqnarray}}
\def\a0{\bar\alpha_0}
\def\thrust{\mbox{T}}
\def\Thrust{\mathrm{\tiny T}}
\def\ae{\alpha_{\mbox{\scriptsize eff}}}
\def\ap{\bar\alpha_p}
\def\as{\alpha_{\mathrm{S}}}
\def\aem{\alpha_{\mathrm{EM}}}
\def\b0{\beta_0}
\def\cN{{\cal N}}
\def\cd{\chi^2/\mbox{d.o.f.}}
\def\Ecm{E_{\mbox{\scriptsize cm}}}
\def\ee{e^+e^-}
\def\enap{\mbox{e}}
\def\eps{\epsilon}
\def\ex{{\mbox{\scriptsize exp}}}
\def\GeV{\mbox{\rm{GeV}}}
\def\half{{\textstyle {1\over2}}}
\def\jet{{\mbox{\scriptsize jet}}}
\def\kij{k^2_{\bot ij}}
\def\kp{k_\perp}
\def\kps{k_\perp^2}
\def\kt{k_\bot}
\def\lms{\Lambda^{(n_{\rm f}=4)}_{\overline{\mathrm{MS}}}}
\def\mI{\mu_{\mathrm{I}}}
\def\mR{\mu_{\mathrm{R}}}
\def\MSbar{\overline{\mathrm{MS}}}
\def\mx{{\mbox{\scriptsize max}}}
\def\NP{{\mathrm{NP}}}
\def\pd{\partial}
\def\pt{{\mbox{\scriptsize pert}}}
\def\pw{{\mbox{\scriptsize pow}}}
\def\so{{\mbox{\scriptsize soft}}}
\def\st{\sigma_{\mbox{\scriptsize tot}}}
\def\ycut{y_{\mathrm{cut}}}
\def\slashchar#1{\setbox0=\hbox{$#1$}           
     \dimen0=\wd0                                 
     \setbox1=\hbox{/} \dimen1=\wd1               
     \ifdim\dimen0>\dimen1                        
        \rlap{\hbox to \dimen0{\hfil/\hfil}}      
        #1                                        
     \else                                        
        \rlap{\hbox to \dimen1{\hfil$#1$\hfil}}   
        /                                         
     \fi}                                         %
\def\etmiss{\slashchar{E}^T}
\def\Meff{M_{\rm eff}}
\def\Ord{\lsim}
\def\OOrd{\gsim}
\def\tq{\tilde q}
\def\tchi{\tilde\chi}
\def\lsp{\tilde\chi_1^0}

\def\gam{\gamma}
\def\ph{\gamma}
\def\be{\begin{equation}}
\def\ee{\end{equation}}
\def\bea{\begin{eqnarray}}
\def\eea{\end{eqnarray}}
\def\lsim{\:\raisebox{-0.5ex}{$\stackrel{\textstyle<}{\sim}$}\:}
\def\gsim{\:\raisebox{-0.5ex}{$\stackrel{\textstyle>}{\sim}$}\:}

\def\ino{\mathaccent"7E} \def\gluino{\ino{g}} \def\mgluino{m_{\gluino}}
\def\sqk{\ino{q}} \def\sup{\ino{u}} \def\sdn{\ino{d}}
\def\chargino{\ino{\omega}} \def\neutralino{\ino{\chi}}
\def\cab{\ensuremath{C_{\alpha\beta}}} \def\proj{\ensuremath{\mathcal P}}
\def\dab{\delta_{\alpha\beta}}
\def\zz{s-M_Z^2+iM_Z\Gamma_Z} \def\zw{s-M_W^2+iM_W\Gamma_W}
\def\prop{\ensuremath{\mathcal G}} \def\ckm{\ensuremath{V_{\rm CKM}^2}}
\def\aem{\alpha_{\rm EM}} \def\stw{s_{2W}} \def\sttw{s_{2W}^2}
\def\nc{N_C} \def\cf{C_F} \def\ca{C_A}
\def\qcd{\textsc{Qcd}} \def\susy{supersymmetric} \def\mssm{\textsc{Mssm}}
\def\slash{/\kern -5pt} \def\stick{\rule[-0.2cm]{0cm}{0.6cm}}
\def\h{\hspace*{-0.3cm}}

\def\ims #1 {\ensuremath{M^2_{[#1]}}}
\def\tw{\tilde \chi^\pm}
\def\tz{\tilde \chi^0}
\def\tf{\tilde f}
\def\tl{\tilde l}
\def\ppb{p\bar{p}}
\def\gl{\tilde{g}}
\def\sq{\tilde{q}}
\def\sqb{{\tilde{q}}^*}
\def\qb{\bar{q}}
\def\sqL{\tilde{q}_{_L}}
\def\sqR{\tilde{q}_{_R}}
\def\ms{m_{\tilde q}}
\def\mg{m_{\tilde g}}
\def\Gs{\Gamma_{\tilde q}}
\def\Gg{\Gamma_{\tilde g}}
\def\md{m_{-}}
\def\eps{\varepsilon}
\def\Ce{C_\eps}
\def\dnq{\frac{d^nq}{(2\pi)^n}}
\def\DR{$\overline{DR}$\,\,}
\def\MS{$\overline{MS}$\,\,}
\def\DRm{\overline{DR}}
\def\MSm{\overline{MS}}
\def\ghat{\hat{g}_s}
\def\shat{\hat{s}}
\def\sihat{\hat{\sigma}}
\def\Li{\text{Li}_2}
\def\bs{\beta_{\sq}}
\def\xs{x_{\sq}}
\def\xsa{x_{1\sq}}
\def\xsb{x_{2\sq}}
\def\bg{\beta_{\gl}}
\def\xg{x_{\gl}}
\def\xga{x_{1\gl}}
\def\xgb{x_{2\gl}}
\def\lsp{\tilde{\chi}_1^0}

\def\gluino{\mathaccent"7E g}
\def\mgluino{m_{\gluino}}
\def\squark{\mathaccent"7E q}
\def\msquark{m_{\mathaccent"7E q}}
\def\M{ \overline{|\mathcal{M}|^2} }
\def\utm{ut-M_a^2M_b^2}
\def\MiLR{M_{i_{L,R}}}
\def\MiRL{M_{i_{R,L}}}
\def\MjLR{M_{j_{L,R}}}
\def\MjRL{M_{j_{R,L}}}
\def\tiLR{t_{i_{L,R}}}
\def\tiRL{t_{i_{R,L}}}
\def\tjLR{t_{j_{L,R}}}
\def\tjRL{t_{j_{R,L}}}
\def\tg{t_{\gluino}}
\def\uiLR{u_{i_{L,R}}}
\def\uiRL{u_{i_{R,L}}}
\def\ujLR{u_{j_{L,R}}}
\def\ujRL{u_{j_{R,L}}}
\def\ug{u_{\gluino}}
\def\utot{u \leftrightarrow t}
\def\ar{\to}
\def\sqk{\mathaccent"7E q}
\def\sup{\mathaccent"7E u}
\def\sdn{\mathaccent"7E d}
\def\chargino{\mathaccent"7E \chi}
\def\neutralino{\mathaccent"7E \chi}
\def\slepton{\mathaccent"7E l}
\def\M{ \overline{|\mathcal{M}|^2} }
\def\cab{\ensuremath{C_{\alpha\beta}}}
\def\ckm{\ensuremath{V_{\rm CKM}^2}}
\def\zz{s-M_Z^2+iM_Z\Gamma_Z}
\def\zw{s-M_W^2+iM_W\Gamma_W}
\def\s22w{s_{2W}^2}

\newcommand{\cpmtwo}    {\mbox{$ {\chi}^{\pm}_{2}                    $}}
\newcommand{\cpmone}    {\mbox{$ {\chi}^{\pm}_{1}                    $}}

\begin{flushright}
{SHEP-10-05}\\
\today
\end{flushright}
\vskip0.1cm\noindent
\begin{center}
{{\Large {\bf The $Z'$ boson of the minimal $B-L$ model \\[0.25cm]
      at future Linear Colliders in $e^+e^- \rightarrow \mu^+\mu^-$}}
\\[1.0cm]
{\large L. Basso, A. Belyaev, S. Moretti and G. M. Pruna}\\[0.30 cm]
{\it  School of Physics and Astronomy, University of Southampton,}\\
{\it  Highfield, Southampton SO17 1BJ, UK.}
}
\\[1.25cm]
\end{center}

\begin{abstract}
{\small
\noindent
We study the capabilities of future electron-positron Linear
Colliders, with centre-of-mass {energy} at the TeV scale, in accessing
the parameter space of   {a $Z'$ boson within  the minimal $B-L$
model}. We carry out a detailed comparison between the discovery
regions mapped over a two-dimensional configuration space ($Z'$ mass
and coupling) at the Large Hadron Collider and possible future Linear
Colliders for the case of di-muon production. As known in the
literature for other $Z'$ models, we confirm that  leptonic machines,
as compared to the CERN hadronic accelerator, display {an additional}
potential in discovering a $Z'$ boson as well as in allowing one
to study its properties at a level of precision well beyond that of
any of the existing colliders.
}

\end{abstract}



\section{Introduction}
\label{Sec:Introduction}
\input{sect_1.tex}


\section{The Model}
\label{Sec:Model}
\input{sect_2.tex}


\section{Results}
\label{Sec:Results}
\input{sect_3.tex}

\input{bibl.tex}
\end{document}

%% file: sect_1.tex

The $B-L$ (baryon number minus lepton number) symmetry plays an
important role in various physics scenarios beyond the Standard Model
(SM). Hence, we consider a minimal $B-L$ low-energy extension of the
SM, consisting of a further $U(1)_{B-L}$ gauge group, three
right-handed neutrinos and an additional Higgs boson generated through
the $U(1)_{B-L}$ symmetry breaking. It is important to note that in
this model the ${B-L}$ breaking can take place at the TeV scale,
i.e. far below that of any Grand Unification Theory (GUT).

In the present proceeding we present some phenomenology related to the
$Z'$ sector of the minimal (no mixing between $Z$ and $Z'$ at
tree-level) $B-L$  extension of the SM at the new generation
of $e^+ e^-$ Linear Colliders (LCs)~\cite{LCs}, considering the
$e^+e^-\to \mu^+\mu^-$ channel as a representative process in order to
study new signatures pertaining to the minimal $B-L$ model.

%% file: sect_2.tex
The model under study is the so-called ``pure'' or ``minimal''
$B-L$ model (see~\cite{B-L:LHC}-\cite{Basso:2009gg} for conventions
and references)
since it has vanishing mixing between the two $U(1)_{Y}$ 
and $U(1)_{B-L}$ groups.
In this model the classical gauge invariant Lagrangian,
obeying the $SU(3)_C\times SU(2)_L\times U(1)_Y\times U(1)_{B-L}$
gauge symmetry, can be decomposed as:
$\mathscr{L} = \mathscr{L}_{YM} + \mathscr{L}_s + \mathscr{L}_f
+ \mathscr{L}_Y$. The non-Abelian field strengths in
$\mathscr{L}_{YM}$ are the same as in the SM whereas the Abelian
ones can be intuitively identified.
In this field basis, the covariant derivative is:
$D_{\mu}\equiv \partial _{\mu} + ig_S
T^{\alpha}G_{\mu}^{\phantom{o}\alpha} + igT^aW_{\mu}^{\phantom{o}a}
+ig_1YB_{\mu} +i(\widetilde{g}Y + g_1'Y_{B-L})B'_{\mu}$.
The ``pure'' or ``minimal'' $B-L$ model is defined by the condition
$\widetilde{g} = 0$, that implies no mixing between the $Z'$ and the
SM-$Z$ gauge bosons.

The fermionic Lagrangian is the usual $SM$ one, apart from the
presence of Right-Handed (RH) neutrinos. The charges are the usual SM
and $B-L$ ones
(in particular, $B-L = 1/3$ for quarks and $-1$ for leptons). The
$B-L$ charge assignments of the fields as well as the introduction of
new fermionic RH-neutrinos ($\nu_R$) and scalar Higgs ($\chi$, charged
$+2$ under $B-L$) fields are designed to eliminate the triangular
$B-L$  gauge anomalies and to ensure the gauge invariance of the
theory, respectively. Therefore, the $B-L$  gauge extension of the SM
group broken at the Electro-Weak (EW) scale does necessarily require
at least one new
scalar field and three new fermionic fields which are charged with
respect to the $B-L$ group.

The scalar Lagrangian is: $\mathscr{L}_s = \left( D^{\mu} H\right)
^{\dagger} D_{\mu}H + \left( D^{\mu} \chi\right) ^{\dagger}
D_{\mu}\chi - V(H,\chi )$, with the scalar potential given by
$V(H,\chi ) = m^2H^{\dagger}H + \mu ^2\mid\chi\mid ^2 + \lambda _1
(H^{\dagger}H)^2 +\lambda _2 \mid\chi\mid ^4 + \lambda _3
H^{\dagger}H\mid\chi\mid ^2$, where $H$ and $\chi$ are the complex
scalar Higgs doublet and singlet fields, respectively.

Finally, the Yukawa interactions are: $\mathscr{L}_Y =
-y^d_{jk}\overline {q_{jL}} d_{kR}H - y^u_{jk}\overline {q_{jL}}
u_{kR}\widetilde H - y^e_{jk}\overline {l_{jL}} e_{kR}H
-y^{\nu}_{jk}\overline {l_{jL}} \nu _{kR}\widetilde H -
y^M_{jk}\overline {(\nu _R)^c_j} \nu _{kR}\chi +  {\rm h.c.}$, where
$\tilde H=i\sigma^2 H^*$ and  $i,j,k$ take the values $1$ to $3$,
where the last term is the Majorana contribution and the others the
usual Dirac ones.

%% file: sect_3.tex
The first thing that we want to explore is the discovery potential of
hadronic and leptonic
machines in the $M_{Z'}$-$g'_1$ plane of our model, in the di-muon
production process. We compare the LHC hadronic scenario
($\sqrt{s}=14$ TeV), with $100\
{\rm fb}^{-1}$ data collected, to two different LC leptonic frameworks, at a
fixed Centre-of-Mass (CM) energy of $\sqrt{s_{e^+e^-}} = 3$ TeV ($500\
{\rm fb}^{-1}$ data altogether) and in a so-called energy scan, where
the CM energy is set to  $\sqrt{s_{e^+e^-}} = M_{Z'} +10$ GeV and we
assume $10\ {\rm fb}^{-1}$ of luminosity for each step. We then limit
both signal and background to the detector acceptance volumes and
$M_{\mu^+\mu^-}$ to an invariant mass window defined by the CMS and
ILC prototype resolution or $3\Gamma_{Z'}$, whichever
the largest (see~\cite{Basso:2009gg}-\cite{Basso:2009hf} for
details). Finally, we define
the significance $\sigma$ as
$s/\sqrt{b}$ ($s$ and $b$ being the signal and background event rates,
respectively): the discovery will be for $\sigma \ge 5$.

As a result, for $M_{Z'} > 800$ GeV, the LC potential to explore the
$M_{Z'}$-$g'_1$ parameter space in the fixed CM energy approach goes
beyond the LHC reach. For example, for $M_{Z'} = 1$ TeV, the LHC can
discover a $Z'$ if $g'_1 \approx 0.007$ while a LC can achieve this for
$g'_1 \approx 0.005$. The difference is even more drastic for larger
$Z'$ masses as one can see from table \ref{mzp-gp-tab}: a LC can
discover a $Z'$ with a $2$ TeV mass for a $g'_1$ coupling which is a
factor $8$ smaller.

\begin{table}[h]
\begin{center}
\begin{tabular}{llll}
\hline
$M_{Z'}$ (TeV)  & \multicolumn{3}{c}{$g_1'$}\\
\hline
  & LHC	& LC ($\sqrt{s}= 3$ TeV)& LC
  ($\sqrt{s}=M_{Z'}+10$ GeV)\\
\hline
 1.0 & 0.0071 & 0.0050 & 0.0026  \\  
 1.5 & 0.011  & 0.0040 &  0.0032 \\ 
 2.0 & 0.018  & 0.0028 &  0.0034 \\ 
 2.5 & 0.028  & 0.0022 &  0.0035 \\ 
\hline
\end{tabular}
\end{center}
\vskip -0.5cm
\caption{Minimum $g_1'$ value accessible at the LHC and a LC for
 selected $M_{Z'}$ values in our $B-L$ model. At the LHC we assume
 $L=100\;{\rm fb}^{-1}$ whereas for a LC we take $L=500\;{\rm
 fb}^{-1}$ at fixed energy and $L=10\;{\rm fb}^{-1}$ in energy
 scanning mode.
\label{mzp-gp-tab}}
\end{table}

In case of the energy scan approach, the $M_{Z'}$-$g'_1$ parameter
space can be probed even further for $M_{Z'} <1.75$ TeV. For example,
for $M_{Z'} =1$ TeV, $g'_1$ couplings can be probed down to the
$2.6\times 10^{-3}$, following a $Z'$ discovery. Furthermore, the
parameter space corresponding to the mass interval $500$ GeV $<M_{Z'}
< 1$ TeV, which the LHC covers better as compared to a LC with fixed
energy, can be accessed well beyond the LHC reach with a LC in energy
scan regime.

Hereafter, we consider the general pattern of the $Z'$ production
cross section in comparison to the SM background as a function of
$M_{Z'}$, for two fixed values of $\sqrt{s_{e^+e^-}}$, in
configurations such that the $Z'$ resonance can be either within or
beyond the LC reach for on-shell production.
The typical enhancement of the signal at the peak is either two
orders of magnitude above the background for $\sqrt{s_{e^+e^-}}=1$ TeV
(ILC configuration) and $g'_{1}>0.05$ or three orders of magnitude
above the background for $\sqrt{s_{e^+e^-}}=3$ TeV (CLIC
configuration) and $g'_{1}>0.1$.
{This enhancement} can onset (depending on the value of $g_1'$, hence
of $\Gamma_{Z'}$) several hundreds of GeV before the resonant mass and
falls sharply as soon as the $Z'$ mass exceeds the collider energy.

While the potential of future LCs in detecting $Z'$ bosons of the $B-L$
model is well established whenever $\sqrt{s_{e^+e^-}} \geq M_{Z'}$, we
would like to remark here upon the fact that, even when
$\sqrt{s_{e^+e^-}} < M_{Z'}$,  there is considerable scope to establish
the presence of the additional gauge boson, through the interference
effects that do arise between the $Z'$ and SM sub-processes ($Z$ and
photon exchange). 

Even when the $Z'$ resonance is beyond the kinematic reach of the
LC, significant deviations are nonetheless  visible in the di-muon
line shape of the $B-L$ scenario considered, with respect to the the
SM case. Incidentally, also notice that such strong interference
effects do not onset in the case of the LHC, because of the smearing
due to the Parton Distribution Functions (PDFs).

Under the assumption that SM di-muon production will be known with a
$1\%$ accuracy we would like to illustrate how the LHC $3\sigma$
observation potential of a heavy  $Z'$ is comparable to a LC indirect
sensitivity to the presence of a $Z'$, even beyond the kinematic reach
of the machine.
This is shown in  table \ref{mzp-gp-tab_ind}, which clearly shows  that
a CLIC type LC will be (indirectly) sensitive to much heavier $Z'$
bosons than the LHC. For example, for $g'_1 = 0.1$, such a machine
would be sensitive to a $Z'$ with mass up to 10 TeV whilst the LHC can
observe a $Z'$ with  mass  below 4 TeV (for the same coupling).
Even a LC with $\sqrt{s_{e^+e^-}} = 1$ TeV (a typical ILC energy) will
be indirectly sensitive to larger $M_{Z'}$ values than the LHC, for
large enough  values of the $g'$ coupling.
For example, such a machine will be sensitive  to  a $Z'$ with mass up
to $7.5$ TeV for $g'_1 = 0.2$ whilst the  LHC would be able to observe
a ${Z'}$ only below $4.7$ TeV or so (again, for the same coupling).

\begin{table}[htb]
\begin{center}
\begin{tabular}{llll}
\hline
$g_1'$  & \multicolumn{3}{c}{$M_{Z'}$ (TeV)}\\
\hline
    & LHC ($3\sigma$ observation) & LC ($\sqrt{s}= 1$ TeV,
    $1\%$ level)& LC ($\sqrt{s}= 3$ TeV, $1\%$ level)\\ 
\hline
 0.05 & 3.4 & 2.2 &  5.5  \\
 0.1  & 4.1 & 3.8 &  10   \\  
 0.2  & 4.7 & 7.5 &  19.5 \\ 
\hline
\end{tabular}
\end{center}
\vskip -0.5cm
\caption{Maximum $M_{Z'}$ value accessible at the LHC and a LC for
 selected $g'_1$ values in the minimal $B-L$ model. At the LHC we
 assume  $L=100\;{\rm fb}^{-1}$.
\label{mzp-gp-tab_ind}}
\end{table}

One interesting possibility opened up by such a strong dependence of
the $e^+e^-\to \mu^+\mu^-$ process in the $B-L$ scenario on
interferences is to see whether this potentially gives unique and
direct access to measuring the $g'_1$ coupling. In fact, notice that 
in the case of $Z'$ studies on or near the resonance (i.e., when
$\sqrt{s_{e^+e^-}}\approx M_{Z'}$), the $B-L$ rates are strongly
dependent on $\Gamma_{Z'}$ (hence on all couplings entering any
possible $Z'$ decay channel, that is, not only $\mu^+\mu^-$). Instead,
when $\sqrt{s_{e^+e^-}}\ll M_{Z'}$ and
$|\sqrt{s_{e^+e^-}}-M_{Z'}|\gg \Gamma_{Z'}$, one may expect that the 
role of the $Z'$ width in such interference effects is minor, the
latter being mainly driven by the strength of $g'_1$. Varying the $Z'$
width as an independent parameter we have proven that the dependence
on $\Gamma_{Z'}$ is negligible. Hence, in presence of a known value
for $M_{Z'}$ (e.g., from a LHC analysis), one could extract $g_1'$
from a fit to the line shape of the cross section at a LC. In fact,
the same method, to access this coupling,  could be exploited at
future LCs independently of LHC inputs, as interference effects of the
same size also appear when $ \sqrt{s_{e^+e^-}} > M_{Z'}$.

%% file: PerugiaProceeding.bbl
\begin{thebibliography}{99}


\bibitem{LCs}
        {K.~Abe {\it et al.}}, [The ACFA Linear Collider Working
        Group], {arXiv:hep-ph/0109166};
        {T.~Abe {\it et al.}}, [The American Linear Collider
        Working Group], {arXiv:hep-ex/0106055},
        {arXiv:hep-ex/0106056}, {arXiv:hep-ex/0106057},
        {arXiv:hep-ex/0106058};
        {E.~Accomando {\it et al.}}, [ECFA/DESY LC Physics Working
        Group], 
        {Phys.\ Rept.} {\bf 299} {(1998)} {1};
        {J.A. Aguilar-Saavedra {\it et al.}}, [The ECFA/DESY LC
        Physics Working Group], {arXiv:hep-ph/0106315}; 
        {K.~Ackermann {\it et al.}}, preprint DESY-PROC-2004-01,
        DESY-04-123, DESY-04-123G.


\bibitem{B-L:LHC}
        {L.~Basso, A.~Belyaev, S.~Moretti and
        C.~H.~Shepherd-Themistocleous},
        {Phys.\ Rev.\  D} {\bf 80} {(2009)} {055030}, [arXiv:0812.4313
        [hep-ph]].

\bibitem{Basso:2009gg}
        {L.~Basso, A.~Belyaev, S.~Moretti, G.~M.~Pruna and
        C.~H.~Shepherd-Themistocleous},
        arXiv:0909.3113 [hep-ph].

\bibitem{Basso:2009hf}
        {L.~Basso, A.~Belyaev, S.~Moretti and G.~M.~Pruna},
        {JHEP} {\bf 0910} {(2009)} {006}, [arXiv:0903.4777 [hep-ph]].




\end{thebibliography}
